\documentclass[12pt,preprint]{aastex}

\begin{document}
\title{Wide Angle X-ray Sky Monitoring for Corroborating
non-Electromagnetic Cosmic Transients}
\author{ Dafne Guetta \altaffilmark{1},\&
David Eichler \altaffilmark{2}
 }

\altaffiltext{1}{Osservatorio astronomico di Roma, v. Frascati 33,
00040 Monte Porzio Catone, Italy; guetta@mporzio.astro.it}
\altaffiltext{2}{Physics Department, Ben-Gurion University,
Beer-Sheva 84105, Israel; eichler@bgumail.bgu.ac.il}

\begin{abstract}
Gravitational waves (GW) can be emitted from coalescing neutron
star (NS) and black hole-neutron star (BH-NS)  binaries, which are thought to be the
sources of short hard gamma ray bursts (SHBs).
The gamma ray fireballs  seem to be beamed into a small solid angle and therefore
only  a fraction of detectable GW events is expected to be observationally
coincident with SHBs. Similarly ultrahigh energy (UHE) neutrino
signals associated with gamma ray bursts (GRBs) could fail to be corroborated by prompt
$\gamma$-ray emission if the latter is beamed in a narrower cone
than the neutrinos. Alternative ways to corroborate
non-electromagnetic signals from coalescing neutron stars are
therefore all the more desirable. It is noted here that the extended X-ray
tails (XRT) of SHBs are similar
to X-ray flashes (XRFs), and  that both can be attributed to an
off-axis line of sight and thus span a larger solid angle than the hard emission. It is proposed that a higher fraction of detectable
GW events may be coincident with XRF/XRT than with hard $\gamma$-rays, thereby
 enhancing the possibility to detect it as
 a GW or neutrino source. Scattered $\gamma$-rays, which
 may subtend a much larger solid angle that the
 primary gamma ray jet,
 are also candidates for corroborating non-electromagnetic signals.

\end{abstract}

\section{Introduction}

Short hard $\gamma$- ray bursts (SHBs) are now suspected to be caused
by the merging of two compact objects, such as neutron stars or
black holes, which would release large amounts of energy over  short
time intervals (e.g. Goodman, 1986).  Collapse of a single object has also
been proposed to give rise to a similar situation (Berezinsky 1987).
Eichler et al (1989) suggested that GRBs could be observed in
coincidence with GW signals when two neutron stars merge.

The huge
isotropic equivalent energy requirements implied by the
BATSE observations of GRB isotropy and submaximal $V/V_{max}$,
suggested that GRBs might be highly collimated (e.g. Levinson and
Eichler, 1993) and this would make them a bad bet to corroborate GW
signals from such mergers, as gravitational radiation is unlikely to
be strongly collimated. This might be fatal (e.g. Guetta and
Stella, 2009) to the original proposal
that LIGO signals would be coincident with GRBs. It is now accepted that GRBs are indeed highly collimated.
 Alternative ways to confirm LIGO signals from
coalescing neutron stars [and, according  some suggestions (Van
van Putten 1999a,b), unstable collapsing disks]  are therefore all the
more desirable.

 The horizons of first generation LIGO and Virgo for NS-NS,
and (BH-NS) mergers are $\sim 20$ and $43$ Mpc,
respectively, while advanced LIGO/Virgo should detect them out to a
distance of $\sim 300$ and $650$ Mpc (for a review see Cutler \&
Thorne 2002). GW signals from NS-NS mergers are expected at a rate
of one in 10-150 years with Virgo and LIGO and one every 1-15 days
with Advanced LIGO/Virgo class interferometers (Berezinsky et al.
2002, 2007). The BH-NS and BH-BH merger rates in the Galaxy are
highly uncertain. Berezinsky et al. (2007) estimate 1\% and 0.1\% of
the NS-NS merger rate, respectively, implying that BH-NS and BH-BH
mergers contribute marginally to the GW event rate, despite the
greater distance out to which they can be detected.

Ultahigh energy (UHE) neutrinos may come from nearby supernova even if an associated
GRB is shaded from our view or entirely smothered by the envelope of the host star (Eichler and
Levinson,
1999). A fluence $F$ of $10^{-4}$erg/cm$^2$ in muon neutrinos at
$10^{12} \le E_{\nu} \le 10^{14}$ eV yields roughly a single
neutrino detection in a gigaton detector such as ICECUBE, the
exact
expectation value depending somewhat upon the energy. An UHE
neutrino signal from a nearby supernova or supernova/GRB could
therefore be detectable at a distance of $D\sim 10^2
E_{iso,\nu,50}^{1/2}$ Mpc. Note that $E_{iso}$ for $\gamma$-rays can
be as high as $10^{54}$ erg [$E_{iso,\gamma,50} = 10^4$]. We face
the following interesting possibility: If the UHE neutrinos from GRB
are beamed into a wider beam than the $\gamma$-rays, then even if
the neutrino efficiency is high, the value of $E_{iso,\nu}$ may be
too low to be seen from any given burst unless it is close.
More importantly, most UHE neutrino events from GRB sources
would not coincide with observed GRBs, as the latter would be most
likely beamed away from us. For corroborating UHE neutrino signals,
as is the case for GW corroboration, we therefore seek
electromagnetic signals that have broader angular reach than the
primary $\gamma$-rays, even at the expense of $E_{iso}$.

We note that several  wide
angle manifestations of nearby GRBs have been proposed.  Eichler and
Levinson (1999) have suggested both high energy neutrino signals and scattered
photons [i.e. scattered off material moving at Lorentz factor less
than the intrinsic opening angle of the primary emission, see also
Eichler and Manis (2007)], each of which could corroborate LIGO
events, at large viewing offsets. Levinson et al (2002) have
suggested orphan afterglows, though there might be some problem
establishing uniqueness via coincidence because of their long
timescales.

 As it
happens, evidence for a high degree of collimation is more
convincing for the long GRBs,  while SHBs are the ones now believed to be
associated with mergers. SHBs frequently show a lower $E_{iso}$, a somewhat broader
$V/V_{max}$ distribution, and less evidence of a narrow opening
angle from jet breaks. This could be understood, for example, if the giant envelope in the case of long bursts provides better collimation than when it is absent.

The presence or absence of the envelope  may be responsible for other differences between short and long GRBs. For example, it may be that there is intrinsic spread in the timescales of the central engines accretion timescale, and that only long bursts are sustained enough to  break through a massive envelope, whereas mergers, perhaps for different reasons, also produce a spread in timescales while allowing all of them to be observed, though this would  explain neither spectral differences nor differences in spectral lags and sub-pulse time scales between short and long GRBs. Neither would it by itself explain why the short duration, hard emission of short GRB lie off to one side of  the  Amati relation while long bursts, X-ray flashes and the X-ray tails of short GRB obey it.

Eichler and Manis (2007, 2008) and Eichler,
Guetta and Manis (EGM, 2009) noted that the unusually hard spectrum
displayed by SHBs, their unusually soft X-ray tail (as compared to the emission of long GRBs), and
their short duration were consistent with a smaller Lorentz factor at the time the short, hard emission is last scattered,
and a larger viewing angle. The larger viewing angle is, {\it a priori},
statistically expected if the line of sight is not obscured by an
extended stellar envelope which is known to exist in the case of
long GRBs, and which would be less likely in the case of NS-NS
mergers. Less collimation and larger allowed viewing offset angle make a coincidence with a
GW signal more likely.  While larger viewing
angle and/or less collimation means smaller $E_{iso}$ and therefore
less $V_{max}$, that is not
 a problem for LIGO collaboration, where the sources
 would in any case be very close.
 \footnote{The suggestion of Eichler and Manis (2007, 2008) and EGM  (2008) that viewing angle affects the perceived durations both of the SHBs hard  emission and X-ray tail is compatible with an additional intrinsic spread in durations of central engine activity for mergers (van Putten  1999b).
The long duration of GRB060614 can be attributed to the prolonged activity of a rotating black hole 
(Van Putten, 2008).  The hypothesis of EGM can also accommodates events such as GRB 060614, which was of long duration while resembling SHBs in other respects. Also, an X-ray tail that lasts $10^2$ s in observer time can result from a SHB whose intrinsic duration is only 1 s. In this paper, however,  we are concerned only with the angular spread  of the X-ray tail, not the intrinsic duration of the central engine activity that causes it, and consider the possibility that the X-ray tails of SHBs may have broader angular spreads and encompass more observers than the short, hard emission.}


Admittedly, the typical viewing angle for SHBs, though perhaps larger than for long GRBs, is uncertain and  could
be small compared to unity.
 There exists by now some evidence that SHBs are beamed, like long GRBs, into a
modest solid angle. Fox et al. (2005)
interpreted the steepening of the optical afterglow light curve of
GRB 050709 and GRB 050724 in terms of a jet break that translates into
a beaming factor $f_b^{-1}\sim 50$ (with $f_b$ the fraction of the
$4\pi$ solid angle within which the GRB is emitted). Soderberg et al
(2006) found a beaming factor of $\sim 130$ for GRB 051221A.
Therefore with the present data the beaming angles of SHB seem to lie
 in the range of  $\sim 0.1-0.2$ radian.

The discovery that X-ray flashes (XRFs) are a class of long GRBs was made by the Wide
Field Camera (WFC) on BeppoSax (Heise et al. 2001). The XRFs are GRBs characterized by no or faint signal in the
$\gamma$-ray energy range, are isotropically distributed in the sky
and have an average duration of $\sim 100$ sec like long GRBs. There
is strong evidence that classical GRBs and XRFs are closely related
phenomena, and understanding what makes them differ could yield
important insights into their origin.  The redshift distribution of
XRFs is very similar to normal GRBs and therefore the high redshift
hypothesis, which might otherwise justify the softness of the burst, cannot
account for all XRFs. D'Alessio et al. (2006) have concluded that the
off-axis hypothesis  seems to be the best hypothesis for now.

Many SHBs show a bright X-ray tail (XRT) that follows the short
prompt $\gamma$-ray emission and lasts for $\sim 100$ s (e.g. Nakar
2007). This X-ray component is evident in SHBs 050709 and 050724,
where the X-ray energy is comparable to or even larger than the energy
in the prompt $\gamma$-rays. It seems that XRTs, though  not detected in all SHBs, are rather common among them. Extrapolation of the late afterglow back to
early times suggests that these tails cannot be interpreted as the
onset of the afterglow emission (Nakar 2007). These X-ray tails have
spectra and durations that are similar to those of the know XRFs,
 and maybe both can be attributed to an off-axis line of sight. In this case, they could encompass more observers than the hard emission of the SHB, and could thus be more opportune for corroborating non-electromagnetic manifestations of mergers and/or core collapses. EGM (2009) made rough estimates of order 0.1 to 0.2 radian offset
from the periphery of the primary fireball, but with large uncertainties.

In this paper we consider the possibility that a wider opening angle of X-ray tails, relative to the hard SHB emission, enhances their likelihood of corroborating non-electromagnetic signals from merger and collapse events.
In  section 2 we report all the properties of the X-ray tails. In section 3
we present a method to determine the XRFs rate. In section 4 we
compare this  rate with the  XRT rate and discuss our results.

\section{X-ray tail properties}

We have considered all
the short bursts detected by Swift from its launch (November 2004)
until August 2009,  this
constitutes a sample of $\sim$ 40 bursts. In Table 1 we report the
observed data.
 We report the properties  of SHB  prompt and X-ray tail emission as
detected by the Swift  X-ray telescope and HETE-2. The X-ray flux is
estimated at 60-100 sec after the burst and is given in the 0.3-10
keV energy range. In the last column we also report what the X-ray
flux would  be if the SHBs were at a distance of LIGO (advanced
version)  detectability (300 Mpc if SHBs come from NS-NS mergers)

\begin{table}[t]
\caption{Properties  of SHBs  prompt and afterglow emission as
detected by Swift  X-ray telescope and HETE-2 (indicated with the *),
 the + indicates that
they could be detected by the WFC. The X-ray flux is estimated at
60-100 sec after the burst and is given in the 0.3-10 keV energy
range. In the last column we report what the X-ray flux would  be if
the SHBs were at a distance of LIGO (advanced version)  detectability
(300 Mpc if SHBs come from NS-NS mergers). }
\begin{tabular}{lllllll}
 \hline
GRB & $z$ & $S_{\gamma}$ & $E_{\gamma, iso}$ & $F_x   $& $ E_{x,iso}$
& $F_x   $ (@ 300 Mpc) \\
   &    & $10^{-7}$ erg/cm$^2$  & $10^{49}$ erg &  $10^{-11}$ erg/cm$^{2}/s$& $10^{49}$ erg
   & $10^{-11}$ erg/cm$^{2}/s$\\
  \hline
 050709*$^+$&0.16  & 3$\pm$ 0.38   &  1.4     & 800 & 3.9  & 3.9 $\times 10^3$ \\
 050724$^+$ & $0.258$    & 6.3 $\pm$ 1    & 7.2        & 1200 & 14.6 &1.4 $\times 10^4$\\
 051210$^+ $&                     & 1.9$\pm$ 0.3 &           & 90 &  & \\
 051221$^+$ & 0.546        & 22.2 $\pm$ 0.8 & 84        & 20  & 0.9 & 923\\
 060313 $^+$&                    & 32.1 $\pm$ 1.4 &         & 30&  &\\
  071227$^+$  & 0.383 &    2.2 $\pm$ 0.3   & 4.0 & 46& 1.1 &  1.1 $\times 10^3$\\
  050509B & 0.225& 0.23$\pm$ 0.09& 0.2 & 0.06 & 5.6$\times 10^{-4}$  &0. 55 \\
 050813 & 0.7& 1.24 $\pm$ 0.46 & 5.2 & 0.6& 0.042 & 42\\
 050906 &      & 0.84 $\pm$ 0.46&   & $<0.007$ &     &\\
 050925 &     & 0.92 $\pm$ 0.18 &    & $ <0.003$  &    & \\
 060502B& 0.287 & 1 $\pm$ 0.13 & 1.15 & 0.1 & 0.001 & 1.4\\
 060801 &       & 0.8 $\pm 0.1$&    & 0.1 &   &\\
 061201 & 0.11 & 3.3 $\pm$ 0.3 & 0.7 & 10 & 0.02& 23\\
 061217 & 0.827 & 0.46 $\pm$ 0.08 & 2.4 & 0.1 & 0.005 &9.1\\
 070429B & 0.904 & 0.63 $\pm 0.1$ &3.5 & 0.11& 0.006 & 10.4\\
 070724 & 0.45 & 0.3 $\pm 0.2$ & 0.6 & 0.05 & 0.0012 &1.7\\
 070729 & 0.904 & 1.0 $\pm$ 0.2&  5.6   & 0.024 &   0.001& 2.5\\
 070809 &             & 1.0                  &        &  0.179 &         & \\
 071112B &                        &  0.48                   &      &$<0.02$&  &\\
 080426 &   & 3.7 $\pm$ 0.3 &  & 0.91 &   &  \\
 080702A &  & 0.36 $\pm$ 0.1 & & 1.0 & & \\
 080905A & &1.4 $\pm$ 0.2 & & 31 & & \\
 081226A& & 0.99 $\pm$ 0.18 & & 0.047 & & \\
 090305A & & 0.75 $\pm$ 0.13 & & 0.55 & & \\
 090426 &2.6 & 1.8 $\pm$ 0.3 &71  & 1.2 & 0.48 & 470 \\
 090621B & & 0.7 $\pm$ 0.1 & & 0.045 & & \\
 \hline
\label{t:fit}
\end{tabular}
\end{table}

In order to give an estimate of the GW events expected from XRFs, XRTs
and SHBs, their cosmic event rates per unit volume and their beaming
factors must be known.
 Measuring the relative detection rates and
distribution of distances of each of the  categories of events
reduces the number of parameters. The universal central engine
hypothesis discussed in EGM (2009), in its
simplest and most naive form, together with the offset viewing
hypothesis for XRFs  posit that SHBs and long GRBs have the same
energetics and that they present XRTs to offset observers the same
way long GRBs display XRFs to such observers.

\section{The rate per unit volume of XRFs in the Local Universe}
\label{rate}
A burst is classified as an XRF when the
Softness Ratio (SR) between the fluences
in the 2-30 KeV band to the 30-400 KeV band is
greater than unity (Lamb \& Graziani 2003).

In this section we give a method to estimate the rate of XRFs (by
which we mean rate per unit volume) following Pelangeon et al. 2008.
This method is valid both for the WFC that works in the 2-28 keV
energy range and for WXM on HETE-2 that works in a similar energy
range. Moreover, the sensitivity of the WFC $\sim 4\times 10^{-9}
{\rm erg cm}^{-2}{\rm s}^{-1}$ (De Pasquale et al. 2006) is
comparable to that of WXM $\sim 9\times 10^{-9} {\rm erg
cm}^{-2}{\rm s}^{-1}  $ (Ricker et al. 2002).

In order to estimate the rate of XRFs  we need to know the redshift of the sources, therefore we select only the XRFs that have determined redshift. These are only the XRFs detected by the WXM as no XRF of
the WFC has a known redshift. Our sample contain 6 long bursts and we report the relevant information about these bursts in table 2.

\begin{table}[t]
\caption{Properties  of the X-ray flashes detected by WXM on HETE-2.  S is the fluence in 2-25 kev}
\begin{tabular}{llll}
 \hline
XRF&  $T_{90}$ & $S$ & $z$  \\
   &   sec &  $10^{-7}$ erg/cm$^2$  & \\
  \hline
 011130   & 39.5  &  5.9   &      0.5        \\
 020317    & 7.14 &   2.2  &       2.11         \\
020903  &  10   &   0.8   &      0.25        \\
030429    & 12.95 &  4.7 &         2.68      \\
030528  & 62.8  &  62    &      0.782        \\
040701 &   11.67&   5.44 &       0.214        \\

 \hline
\label{t:fit}
\end{tabular}
\end{table}

The steps to compute the rate are:

\begin{itemize}
\item
Determine the maximum redshift $z_{max}$ at which the source, an XRF,  could have been
detected by the instruments, by first comparing the
measured flux with the threshold flux of the instrument for an XRF with known distance (z):
$F_x/F_T=(D_{\rm max}(z_{\rm max})/D(z))^2$

\item assume that the GRB rate follows the star formation rate.
For this we have adopted the model SFR$_2$ of Porciani \& Madau (2001)
that reproduces a fast evolution between $z=0$ and $z=1$
and remains constant beyond $z\ge2$

\begin{equation}
R_\mathrm{SFR_2}(z)\propto\,0.15 h_{65} \frac{exp(3.4z)}{exp(3.4z)+22}
\end{equation}

\item derive for each burst the number of GRBs per year within its visibility volume
\begin{equation}
\label{NV}
N_{\rm Vmax}\propto\,\int_0^{z_{\rm max}} dz\,\frac{dV(z)}{dz}\,\frac{R_{SFR_2}(z)}{1+z}
\end{equation}
In this equation $dV(z)/dz$ is the comoving element volume, described by
\begin{equation}
\frac{dV(z)}{dz}=\frac{c}{H_0}\,\frac{4\pi\,dl^2(z)}{(1+z)^2\,[\Omega_{M}(1+z)^3+\Omega_{K}(1+z)^2+\Omega_{\Lambda}]^{1/2}}
\end{equation}
where $H_0$ is the Hubble constant, $\Omega_{K}$ is the curvature contribution
to the present density parameter ($\Omega_{K}=$~1$-\Omega_{M}-\Omega_{\Lambda}$),
$\Omega_{M}$ is the matter density and $\Omega_{\Lambda}$ is the vacuum density.
Throughout this paper we have assumed a flat $\mathrm{\Lambda}$CDM universe where
($H_0$, $\Omega_{M}$, $\Omega_{\Lambda}$)$=$(65$h_{65}$~km~s$^{-1}$Mpc$^{-1}$, 0.3, 0.7).\\
This procedure allows us to give each burst a weight ($W_b$)
inversely proportional to $N_{Vmax}$. The rationale of weighting
each burst by $1/N_{Vmax}$ is the following: the visibility volume
is different for each XRF of our sample. Moreover, each XRF observed
is randomly taken from all the bursters present in its visibility
volume. In this way, rare bright XRFs, having a large visibility
volume, will have low weights, while faint local XRFs will have
higher weights. This procedure also takes into account the fact that
the XRF rate evolves with redshift, leading to the fact that XRFs are
about ten times more frequent at $z\sim1$ than at present.
\end{itemize}

This study also allows us to derive the rate of XRFs detected
and localized by HETE-2 ($R_0^\mathrm{H2}$).
For that purpose, we consider that each XRF in our sample contributes
to the local rate in proportion to:
\begin{equation}
h_b=\frac{N_{Vloc}(z=0.1)}{N_{Vmax}(z=z_{b,max})}
\end{equation}
with the $N_V(z)$ computed according to Eq.~\ref{NV}, and we obtain
the number per unit volume of HETE-2 GRBs during the mission,
\begin{equation}
\tau=\frac{1}{V_{loc}}\sum_{b=1}^{n_{burst}}h_b
\end{equation}
the value of which we calculated to be
~ 7.7 Gpc$^{-3}$, similar to what found by Pelangeon et al. 2008.\\
\\
In order to normalize this in terms of  rate per year,
we took into account the effective monitoring time of the WXM, obtained from:
\begin{equation}
T_{m} = \frac{T_\mathrm{mission} \times T_\mathrm{\epsilon}}{4\pi} \times S_\mathrm{cov}
\end{equation}
where $T_m$ is the effective monitoring time of the WXM, $T_{mission}=$~69~months
is the duration of the mission, $T_{\epsilon}=$~37\% is the mean observation
efficiency during $T_{mission}$, and $S_{cov}=2\pi(1-cos~45\degr)=1.84$~sr
the sky-coverage of the WXM\footnote{Recall that throughout this study,
we only consider the XRF localized by the WXM.}.\\
The effective monitoring time of the WXM is hence $T_{m}=$~0.31~yr.
Using this, the rate of detectable XRFs in the Local Universe per
Gpc$^3$ and per year can be found to be
$\sim$25~Gpc$^{-3}$yr$^{-1}$. This result is a lower limit because
HETE-2 missed the bursts with intrinsic peak-energy $E_p$  lower than 1
or 2~keV as well as bursts
occuring at very high redshifts (Pelangeon et al. 2008).\\
Note that the main contribution to the rate comes from XRF 020903,
which has a maximum detectability redshift $z_{max}\sim 0.3$,
 implying a small visibility volume and therefore a large
weight.

\subsection{Swift XRFs}

Because the Swift BAT instrument , which provide the trigger conditions, has an energy band
(15-150 kev) narrower than WXM+Fregate on HETE2, we have to find another way to define an XRF.
Sakamoto et al. (2008) define an XRF to be a burst with the ratio of the fluence in the 25-50 KeV band to that in the 50-100 KeV band  greater than 1.32 and we use this definition to construct a sample
of XRF with redshift.
The relevant properties are given in table 3

\begin{table}[t]
\caption{Properties  of the X-ray flashes detected by Swift, S is the fluence in 15-150 kev and F is the 1-sec Peak Photon Flux}
\begin{tabular}{lllll}
 \hline
XRF&  $T_{90}$ & S& $z$  & F\\
   &   sec &  $10^{-7}$ erg/cm$^2$  &  &ph/cm$^2$/s\\
  \hline
 050315  &  95.6 &    32.2 &    1.949 &     1.93       \\
050319  &  152   &  13.10  &     3.24 &   1.52        \\
050406  &  6.4   &  0.79     &   2.44   &       0.36      \\
050416   &  3.0  &   3.7      &   0.6535  &         4.88    \\
050824   & 26.6  &  2.7     &    0.83    &       0.5     \\
051016   & 4.0    & 1.7      &   0.9364 &        1.3       \\
060108    &14.3   & 3.69   &     2.08  &         0.77       \\
060218    & 2100  &  15.7 &       0.033&       0.25  \\
060512    & 9.7     & 2.3     &    0.44     &         0.88  \\
 \hline
\label{t:fit}
\end{tabular}
\end{table}

In order to estimate the rate of XRFs from
the Swift data we can repeat the analysis described above taking the
threshold of Swift in 15-150 kev of $\sim $ 0.25 ph/cm$^2$/s (Band
2006). We then consider 4.5 years of activity with a sky coverage of
0.17 sr and find a rate of R$\sim$ 130 Gpc$^{-3}$ yr$^{-1}$. Note again
that the main contribution to the rate comes from XRF 060218, which 
has a maximum redshift $z_{max}\sim 0.03$ implying a small
visibility volume and therefore a  large weight. This rate
determined by XRF 060218 is similar to what found in Guetta \& Della
Valle (2007). The soft $\gamma$-rays of XRF 060218, which show a
spectral evolution similar to many other GRBs and subpulses therein,
may be  photons scattered off relatively slow ambient material
(Mandal and Eichler, in preparation).

Comparing this rate and the rate found with the XRFs of the WXM
 with the rate found by Guetta et al. (2005) and Guetta \& Piran (2007)
for long GRBs,  $\sim$ 0.1-0.4 ~Gpc$^{-3}$yr$^{-1}$,
we infer that the population of $\gamma$-ray bursts
is dominated by the X-ray flashes at $z<0.1$. This is understandable, as XRFs
are soft but also faint in the observer frame, according to the
hardness-intensity relation derived by Barraud et al.
(2003). Therefore, if the rate of detected bursts is in
fact higher for classical GRBs than for XRFs, we can guess that this
is because the classical GRBs can be seen out to greater distances.

That XRFs have a much higher rate per unit volume than classical GRBs,
within the framework of our hypothesis that they are the same
phenomenon viewed from different angles, suggests that  the
opening angle of XRF is significantly wider than for the GRB.
Significantly, the supernova-associated GRB and XRF event rate is
much larger not only than the classical GRB rate (Guetta \& Della
Valle 2007) but also larger than the estimated rate of NS merging.
The main contribution to the supernova-associated GRB event rate comes  from GRB 980425 at z=0.0085.  This burst was detected by the
{\it Beppo}SAX WFC (Pian et al. 1999) and by BATSE (Kippen 1998).
The peak flux in the 50-300 kev band was $F_{50-300}=4.48$ ph cm$^{-2}$s$^{-1}$.
Given the threshold of BATSE, $\sim 0.25 $ ph/cm$^2$/s  we find that D$_{\rm max}$ =160 Mpc.
The {\it Beppo}SAX sky coverage was about 0.08 and the operation
time $\sim$ 4 years. Therefore the rate of 980425-like events is $R\sim 182$ Gpc$^{-3}$yr$^{-1}$
which is $\sim $ 10 times higher than the XRFs rate and more than 100 times higher than
the high luminous "classical" GRB rate.
This high rate suggests that if classical GRB emit GW, e.g. from an
unstable protocollapsar tori (van Putten 1999a,b) then combined signals
from wide angle electromagnetic emission and GW might be the most
common sort of electromagnetic plus non-electromagnetic multi-detections of mergers/collapses.

\section{Rate of XRTs and discussion}

The rate of XRFs is an upper limit to the rate of XRTs. For the lower limit
we can take the one of SHBs derived by Guetta and Stella (2009).
  In this paper they find evidence in
favor of a bimodal origin of SHB progenitors where a fraction of SHBs
comes from the merging of primordial neutron star-neutron star (black hole)
and a fraction comes from the merging of dynamically formed
binaries in galaxy clusters.
In particular they find that  the redshift distribution of SHBs
is best fitted when the  incidences of primordial and dynamical
mergers among the SHB population are 40\% and 60\% respectively.
In this case the rate of SHB is
$R_0\sim 2.9$~Gpc$^{-3}$yr$^{-1}$.

For a fiducial value of $f_b^{-1}\sim 100$, we derive a beaming-corrected rate of
$\rho_0 =f_b^{-1} R_0\sim 290(f_b^{-1}/100)$.
Therefore the rate of XRTs is $2.9<R<130$ Gpc$^{-3} $yr$^{-1}$. This rate is left
uncorrected for the beaming as we don't know the beaming angle of this X-ray emission
which can be the same or larger than the beaming angle of the $\gamma$-ray emission.

Another way to estimate the rate of the XRTs is to consider the X-ray tails
detected by the Swift X-ray telescope that could be detected by the WFC
if they were at a distance of 300 Mpc.  These are four
 GRBs (GRB 050709, GRB 050724, GRB 051221, GRB 071227).
 Considering the threshold of the detector that triggered them (Fregate for 050709
 and BAT for the other three bursts)  and
using the same procedure described above for the XRFs rate, we find a
rate of $\sim 7$ Gpc$^{-3}$yr$^{-1}$.

Our suggestion that some XRTs  of SHBs are XRFs,
 combined with
 the hypothesis that they correspond to offset viewing of
 a long burst in some
other direction (Eichler, Guetta and Manis, 2009)  predicts that a
large enough sample of XRFs, even if unbiased by any $\gamma$-ray
trigger, should have a subset that correlates with SHBs. A careful
analysis, however, shows that BATSE should have detected less than
one SHB coincident with any XRF recorded by other contemporaneous
detectors. A larger sample of XRFs detected while a SHB detector is
operating would give tighter constraints.


In summary, we  have considered the possibility  that SHBs have
larger opening angles than long GRBs, and that the XRT associated
with the SHB has a wider angular extent than the harder emission.
Very rough estimates of the opening angle of XRTs, based on their
hypothesized similarity to XRFs, is an opening angle of 0.1 to 0.2
radians [EGM 2009], which may be somewhat larger than estimates of
the opening angles for the hard emission, which are typically 0.03
to 0.1 radians (Bloom et al. 2003).
 While this does not constitute proof that the XRTs have
 larger opening angles than the hard $\gamma$-ray emission from the SHBs,
the fact that extended soft emission is a reliable indicator for
SHBs (Donaghy  al. 2006) suggests that the solid angle in which the
soft photons are detectable by HETE II  is at least as large as that
from which the hard $\gamma$-beam is detectable. { On the other
hand, our estimate for the rate per unit volume of XRTs that would
be visible from the typical distance of an advanced LIGO source,
$\sim 300 Mpc$, is  less than the estimated rate of neutron star
mergers, which  leaves open the possibility that even the XRTs are
beamed and could not corroborate most LIGO signals. Further
information on the relative detectabilities of XRTs and the
corresponding short hard $\gamma$-ray emission could be obtained by a wide
field X-ray camera and $\gamma$-ray detectors working together.

In any case, the rate of WFC-detectable XRTs per sphere of radius
300$R_{300}$ Mpc is at least about 0.8$R_{300}^3$ per year, meaning
that a $2\pi$ detector would detect one per 2.5 years for
$R_{300}=1$. This is a non-negligible if modest fraction of the
total expected rate
 of LIGO
signals from mergers,  about 3 per year with advanced LIGO.
Including the other two WFC-detectable XRTs, though their redshift is unknown,
would raise the estimate
to about 1 $\times R_{300}^3$ per year. This suggests that some fraction of LIGO
signals, if not most or all, could be corroborated by wide field
X-ray cameras.

Coincidentally, this rate of 1 per several years is about the rate
of supernova-associated GRB within 300 Mpc, as estimated from the
prototypes GRBs 980425 and 060218, which could be looked for in
coincidence with UHE neutrinos. We also find the event rate per unit
volume for supernova-associated GRBs and XRFs to be about $10^2$
higher  than for cosmologically distant GRBs. If gravitational waves
and/or neutrinos are emitted by such events, then nearby
SN-associated GRBs, corroborated by wide angle EM emission such as
XRFs or scattered $\gamma$-rays, may be the most frequent collapse
events observed simultaneously in both electromagnetic and
non-electromagnetic modes.

 We acknowledge support
from the U.S.-Israel Binational Science Foundation, the Israel
Academy of Science, and the Robert and Joan Arnow Chair of
Theoretical Astrophysics, and in part from the National Science Foundation under grant number NSF PHY05-51164.

\end{document}